# Effect of processing parameters on the properties of freeze-cast Ni wick with gradient porosity

P.J. Lloreda-Jurado, E. Chicardi, A. Paúl, R. Sepúlveda*

Departamento de Ingeniería y Ciencia de los Materiales y del Transporte, E.T.S. de Ingenieros, Universidad de Sevilla, Avda. Camino de los Descubrimientos s/n., 41092 Sevilla
*Correspondence: rsepulveda@us.es

**Abstract:** Wicks are the main component of Loop Heat Pipe systems, whereby coolant liquid flow through their porous structure. They are usually formed by a primary wick to produce liquid transportation by capillary force, and a secondary wick that is continuously wetted by the liquid coolant. Traditionally, the two wicks are manufactured separately and subsequently joined, thereby creating an interface that reduces the liquid transfer efficiency. In order to overcome this situation, a gradient porous wick is proposed and successfully manufactured through the freeze-casting method in a single operation. The influence of two different dispersant agents, KD4® and stearic acid was studied on the processing parameters, final pore size and morphology, and capillarity performances. A variate of gradient porosity was obtained by applying a diverse thermal gradient and solidification front velocity during directional solidification. The rheological characterisation of the camphene-based NiO suspensions was performed using a rotational viscometer. The final pore size and morphology were characterised by Optical Microscopy, Field Emission Scanning Electron Microscopy, and X-ray computed tomography. The use of stearic acid improves the particle stabilisation and generates pore enlargement with an equiaxed pore structure, while commercial dispersant KD4® shows a dendritic pore morphology at lower thermal gradient.

**Keywords:** Freeze casting, Nickel, wick, camphene, stearic acid, nanoparticles

## 1. Introduction

Porous metallic materials are used in many applications, such as filters [1], bone replacement implants [2], metal-supported solid oxide fuel cells [3], and wicks for loop heat pipe (LHP) systems [4]. Nowadays, LHPs play a major role in improving the performance of microelectronic devices [5,6], spacecraft components [7], and solar water-heating systems [8]. An LHP system is an efficient heat-transfer device that works on a closed condensation-evaporation loop of a cooling fluid. Figure 1 shows a schema of the LHP mechanisms, wherein the capillarity forces produced by a porous wick eliminate the need for any pumping equipment. Wicks are manufactured by means of loose sintering or cold-pressing with space holders and by using powdered materials such as nickel [9,10], titanium [11], stainless steel [12], or copper [13,14] due to their high corrosion resistance and low-to-medium thermal conductivity. To ensure high heat-transfer capability during continuous operation, wicks should have high permeability and good capillarity with a low pressure re drop. These features are obtained with an open pore structure with pore size and porosity ranges of 1-100 ⍴m and 50-75%, respectively [15,16]. Additionally, sufficient mechanical strength (~10 MPa) must be guaranteed for dimensional tolerance adjustment, usually performed by electrical discharge machining. Since wicks need to be continuously wetted by the cooling fluid, a two-wick system is employed [17–19]. A primary wick with a smaller pore size ensures transportation of the liquid to the evaporator chamber, while a secondary wick with large pore size or of a different material connects the primary wick with the compensation chamber, thereby maintaining the condenser unit filled with the cooling fluid.

Nevertheless, the two parts are manufactured separately and are subsequently joined, which creates an interface that could reduce the efficiency of the fluid transfer.

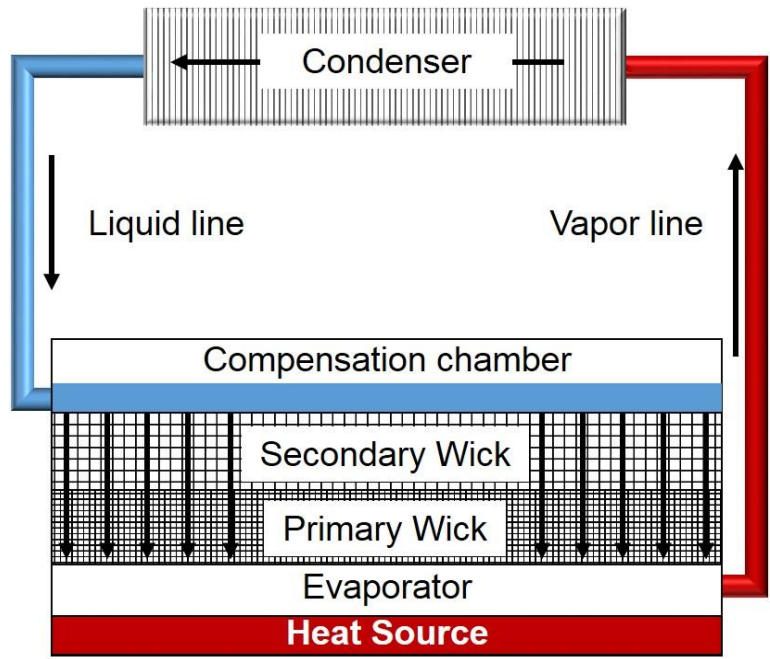

Figure 1. Schema of a loop heat pipe

The efficiency of LHPs is linked to their ability to extract heat, which is reduced in electronic applications since wicks become thinner and smaller. To counterbalance this effect, materials of medium-range thermal conductivity are being tested with new designs of porous structures. In this respect, a new type of wick with a biporous capillary structure has been receiving much attention [9,10,20] due to its resulting increase in the heat-transfer coefficient in the evaporation zone. According to Maydanik et al. [15], this pore arrangement enables the effective evaporating layer of the wick to be increased, and the evaporator thermal resistance to be reduced by diminishing the average thickness of the walls.

For the development purposes of a new wick showing a gradient pore structure that eliminates the joining process, the Freeze-Casting (FC) technique could present a suitable and interesting option. Freeze casting is a simple, low-cost and scalable method [21], which is appropriate for the creation of porous materials with tailored pore features, simply by altering the processing parameters [22,23]. This technique relies on the solidification of a liquid phase in which particles have been dispersed 1since dendrites grow during solidification and push and concentrate the suspended particles into the interdendritic spaces. The sublimation of the solidified phase creates a green body powder scaffold, whose struts or walls are sintered at elevated temperature. The final porous structure can be adjusted through the selection of the raw powder material, the liquid medium [23], and the solidification rate. This technique has been extensively employed with submicronmetric ceramic powders using water [24] or camphene [25] as the dispersant media. Recently, the use of copper [26,27], nickel [28], and iron [29] oxide particles suspended in camphene has been successful in producing highly interconnected metallic porous structures, with wide pore-size distribution [30].

The use of nickel oxide particles through the FC process in water-based suspensions is widespread since it obtains porous materials for the application of solid oxide fuel cells (SOFCs) [31,32], while the use of camphene-based suspensions to produce Ni wicks remains limited. Nonetheless, recent research successfully developed Ni wicks by FC using a water-based suspension of metal particles [33]. The employment of camphene offers a wide range of benefits, such as anisotropic branched morphology with gradual pore sizes, large and elongated pores in the direction of solidification, and high percentages of interconnected porosity. Camphene also has a melting point above room temperature (42-44 ºC) and it can be sublimated at ambient conditions, which reduces energy consumption during solidification and facilitates the sublimation process. However, its non-polar characteristic narrows the options for suitable additives. Therefore, most researchers have used polystyrene (PS) and anionic polyester oligomeric (commercially known as KD4® or Zeprhym®) as a binder and dispersant agent, respectively [23,34,35].

The main goal of this research involves the manufacture of a suitable Ni wick with interconnected gradient pore-size structure through FC technology, by employing NiO nanoparticles as the raw material. Furthermore, the influence of two dispersant agents, stearic acid and KD4, on the rheological behaviour of slurries is studied, together with the final pore morphology and the capillary performance of Ni wicks. Finally, the pore-size gradient is correlated as a function of the cooling conditions applied.

## 2. Experimental Procedures

*2.1 Wick manufacturing*.

Two different camphene (95% purity, Sigma Aldrich, Madrid, Spain) suspensions with 5 vol.% of NiO nanoparticles (20-30 nm in diameter, GNM Oocap France SAS, St-Cannat, France) were created by adding 8 wt.% of Hypermer™, KD4®, (CRODA Ibérica SA, Barcelona, Spain) or stearic acid, SA, (98% purity, Alfa Aesar, Barcelona, Spain) as the dispersant agent. Each suspension was mixed by ball milling at 60 ºC, where the dispersant was first incorporated into melted camphene and then mixed for 30 min. The NiO was then added and milled for 8 h. Finally, 20 vol.% of PS (Mw=350,000 g/mol, Sigma Aldrich, Madrid, Spain) was incorporated as the binder and mixed for an extra 3 h. The dispersant and binder doses were in accordance with the powder load.

After milling, the suspension was immediately poured into a PTFE mould with a Cu base, preheated at 60 ºC inside an incubator and pre-rested for 15min before the solidification process started. An assisted solidification process, reported elsewhere [30], was initiated by running water at 30, 35, 40, or 42.5 ºC through the mould base, and by reducing the incubator temperature by 0.2 ºC/min. Green samples measured approximately 30 mm in diameter by 16 mm in height.

Sublimation of solid camphene within the samples was completed after 48 h in ambient conditions. Green samples underwent heat treatment at 600 ºC for 2 h for organic burn-out in a $N_2$ gas flow (Airliquide, Seville, Spain), followed by a sintering process under reducing conditions at 1100 ºC for 3 h with Ar-20$H_2$ gas flow (Airliquide, Seville, Spain). Heating and cooling rates were set at 1 ºC/min and 5 ºC/min, respectively, to prevent cracks and sample distortions. Wick samples were denoted according to each cooling temperature and to the dispersant agent used: for example, 30SA corresponds to a sample solidified at 30 ºC and using stearic acid as the dispersant agent.

*2.2 Wick characterisation.*

In order to determine the influence of the dispersant agent and the milling time on the suspension viscosity, a Myr VR 300 rotational viscometer with a low-viscosity adaptor (Viscotech Hispania S.L.) was employed to measure the viscosity of each suspension at 4, 8, 12, and 24 h of milling times. Suspensions were first sonicated for 15 min and then kept inside the viscometer chamber for another 15 min to reduce the presence of air bubbles. All measurements were performed at 60 ºC. At each milling time, the viscosity was determined by sequentially increasing the shear rate from 36.45/s up to approximately 121.5/s. Between measurements, an interval of 2 min was fixed to stabilise the dispersion system and prevent the formation of vortices due to the increasing shear rate. The viscosity values were kept constant regardless of whether they had been measured by increasing or decreasing the shear rate, and therefore no hysteresis loop was observed and the appearance of thixotropy or rheopexy effects could be considered negligible.

Two processing FC parameters, the solidification front velocity ($V^h$) and the average thermal gradient ($G^h$), were monitored by 4 thermocouples placed along the sample height ($h$) at 0, 4, 8, and 12 mm. $V^h$ was calculated as the time that the suspension solidification temperature ($T_{ss}$) takes to move between thermocouples (4 mm), and $G^h$ was defined as the temperature difference between sequential thermocouples at the moment that $T_{ss}$ reached halfway. The following equations show the calculation procedure in greater detail.

$$V^h = \frac{0.4\ cm}{t_{ss}^h - t_{ss}^{h-4}}\ \left[\frac{cm}{s}\right], (h = 4, 8, 12\ mm) \qquad (1)$$

$$G^h = \frac{T_\tau^h - T_\tau^{h-4}}{0.4\ cm}\ \left[\frac{°C}{cm}\right]\ with\ \tau = (t_{ss}^h - t_{ss}^{h-4})/2\ ,\ (h = 4, 8, 12\ mm) \quad (2)$$

where $t_\tau^h$ is the time to reach $T_{ss}$ at $h$, and $T_\tau^h$ is the temperature measured at $h$ at the time $\tau$. $T_{ss}$ was obtained by differential scanning calorimetry (DSC) analysis at 43.3 ºC.

The average interconnected porosity was determined through Archimedes' method, which involved soaking the wick samples in hot water for 24 h. Several Optical Microscopy (OM) and Field Emission Scanning Electron Microscopy (FESEM) images were taken on the central axial plane at different heights of each fabricated sample. These images were employed to estimate the average pore size and total porosity across the sample height by using the non-redundant maximum-sphere-fitted image analysis technique [36]. The pore morphology and interconnectivity of samples 30KD4 and 42.5KD4 were determined by X-ray computed tomography (X-CT) for comparison. X-CT measurements were carried out with a voxel of 3 μm on sub-samples (with an approximate volume of 1.2 mm x 1.2 mm x 2.5 mm) located in the lower and upper parts. The collected 2D cross-section images were pre-processed with ImageJ software (8-bit binarisation, threshold, and removal of 1-pixel outliers) and 3D visualisations were made using Avizo software.

In order to determine the capillary performance (i.e., the amount of liquid absorbed over time), samples were carefully placed in contact with the water reservoir by means of a micrometric screw. The weight of water adsorbed over time was ascertained using electronic scales. The top sample surface was placed in direct contact with the water to reproduce the effect of impregnation of the secondary wick and the capillarity suction towards the primary wick. Figure 2 shows the experimental setup used in this work with details of the sample orientation (Figure 2a) and the holding-movement system (Figure 2b). Similar setups can be found in the literature that use water as the adsorbed liquid [37–39].

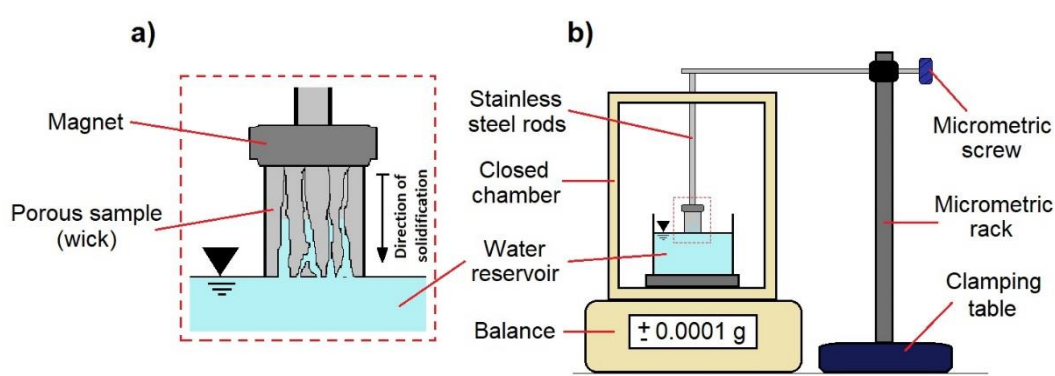

Figure 2. Experimental setup employed to determine the capillary performance on the Ni wicks. Details of the sample orientation (a) and the holding-movement system (b) are shown

**3. Results and Discussion**

3.1. Rheological behaviour of NiO camphene suspensions

Figure 3 shows the influence of the two different dispersant agents, KD4® and SA, on the rheological behaviour of 5 vol.% NiO camphene-based suspension over the milling time. These results illustrate the effect of the carbon-chain length when NiO nanoparticles are used. KD4 is a dispersant agent widely used in camphene suspensions prepared for freeze casting, since it achieves suitable dispersion over submicron and micrometre particles. According to Moloney et al. [40], KD4 and SA employ their carboxyl functional group as a particle anchor. This functional group allows the formation of chelate complexes around the particle surfaces, keeping it strongly adsorbed to the particle while the other side of the molecule shows good affinity with the solvent. Both dispersants reduce the particle attraction forces through the steric barrier created by their carbon chains. The main difference between these two dispersants is the carbon-chain length of 10 and 4 nm for KD4 and SA, respectively. In the case of submicronmetric particles, only KD4 is capable of achieving complete steric stabilisation, while SA could cover and isolate the particle successfully but its chain length is insufficient to achieve a complete steric stabilisation [41].

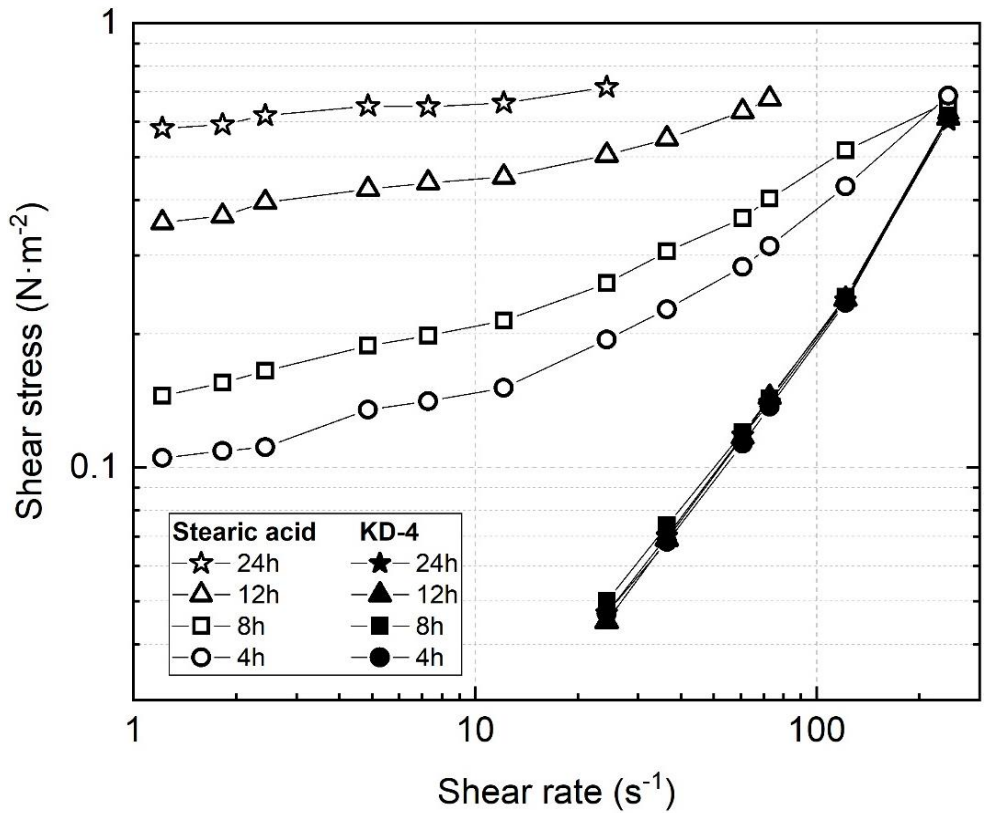

Figure 3. Influence of the KD4 (solid marks) and SA (hollow marks) dispersant agents on the rheological behaviour of 5 vol.% NiO camphene-based suspensions over the milling time (4, 8, 12, and 24 h).

In the case of KD4®, Figure 3 indicates that the rheological behaviour does not change significantly with the milling time. A calculated viscosity of 2.32 mPa·s was determined using a power-law model for a near-Newtonian fluid. This low viscosity value could be attributed to the low attractive energy existing between the particles due to the shielding effect produced by the long polymeric chain absorbed. The long chains provide a thick layer of isolation, which reduces the electrical potential in the diffuse layer and results in a strong repulsion between particles due to the steric effect [42].

When SA is used, the suspension presents rheological behaviour as a non-Newtonian fluid and the shear stress increases with the milling time, as can be deduced from Figure 3. Moreover, suspensions using SA show viscoelastic behaviour that can be fitted to a Herschel-Bulkley rheological model. Table 1 shows the parameters obtained, such as yield point ($\tau_0$), viscosity ($\eta$), and flow index ($n$) at various milling times. As the milling progresses, the yield point and the viscosity increase significantly and the flow index is reduced. The flow index ($n < 1$) obtained denotes shear-thinning behaviour of the NiO camphene-based suspension. The increase in viscosity with the milling time could be attributed to the formation of a mono-disperse particle state, which is induced by the shear forces during the milling [43,44]. The particle agglomerates in the suspension break up and hence each new particle surface could be covered by SA. Therefore, the viscosity increases as a homogenous smaller particle size in the suspension is obtained.

Table 1. Herschel-Bulkley model parameters at different milling times

Herschel-Bulkley model: $\tau = \tau_0 + \eta \cdot \gamma^n$

| Milling time(h) | $\tau_0$ (N·m$^{-2}$) | $\eta$ (mPa·s) | $n$ |
|---|---|---|---|
| 4 | 0.10 | 6 | 0.83 |
| 8 | 0.13 | 17 | 0.64 |
| 12 | 0.34 | 27 | 0.58 |
| 24 | 0.58 | 27 | 0.54 |

This shear-thinning behaviour is an indication of weakly attractive, but non-touching, particle interaction. This weakly attractive interaction 1leads to a high-packing factor during particle aggregation 1throughout the process of solidification [44,45]. Therefore, SA constitutes a promising dispersant alternative in camphene-based suspensions of nanoparticles for its use in freeze casting.

3.2 Microstructures of manufactured wicks

The differences in the particle interaction phenomena for each dispersant, previously verified by rheology characterisation, exert a major effect over the microstructure parameters of the wicks manufactured, such as the amount and type of porosity, the pore size, and its morphology. Figure 4 shows the effect of the dispersant agent and the cooling temperature used on the calculated porosity across the sample height. KD4 creates a gradient porosity at all cooling temperatures, with the lowest porosity obtained in the lower part of the sample, while SA generates a practically constant value of porosity. Furthermore, the total porosity seems higher with the use of SA. This feature is more evident at the lower end since sample 30KD4 reaches 42% porosity and 30SA climbs up to 84% (Figure 4). The porosity reduction in the lowest parts of those samples that use KD4 in their suspensions could be ascribed to particle sedimentation during the pre-resting and solidification steps.

This sedimentation modifies the particle concentration across sample height. As a consequence, the concentration in the liquid above the solidification front is continuously reduced as solidification progresses. This event results in a gradient total porosity, which varies from 42% at the lower part of the 30KD4 sample to 90% in the upper part of the 42.5KD4. Furthermore, the densification in the lower zone of samples dispersed with KD4 is accentuated when the cooling temperature employed is lower (i.e., when the solidification front velocity ($V^h$) and the average thermal gradient ($G^h$) are higher), and increases from 42%, in samples 30KD4 (cooled at 30 ºC), up to 63% in samples 42.5KD4 (cooled slowly at 42.5 ºC). Particle sedimentation could be attributed to the formation of heavier particles due to various factors, such as particle agglomerates not fully dispersed during the milling process, the higher molecular weight of KD4 (1414.3 g/mol) attributed to its long carbon chains, and a certain level of entanglement occurring between the PS and camphene molecules [40,46]. Likewise, this effective enlargement of the particle size decreases the critical velocity for particle engulfment ($V_{cr}$) [47]; at lower $V^h$, particles will be surrounded by the growing camphene crystal during solidification, instead of being pushed into the interdendritic spaces. Therefore, a pore morphology with smaller pore size could be produced.

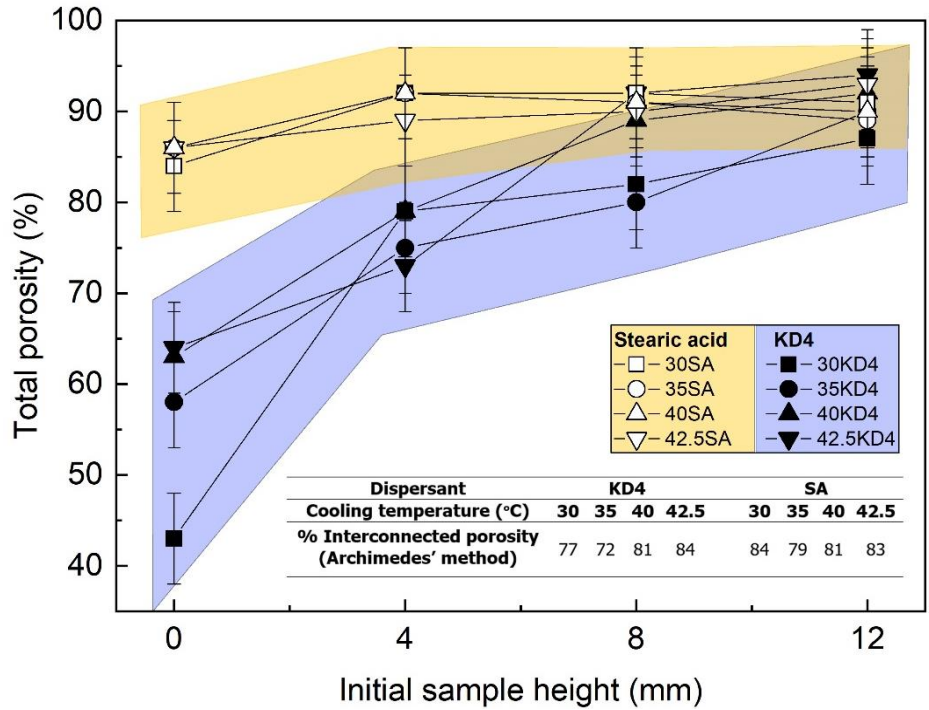

Figure 4. Influence of the dispersant agent and the cooling temperature on the total porosity of the Ni wick across the sample height. Total sample porosity was calculated using image analysis over the optical microscopy images taken at the corresponding initial sample height. The table inserted shows the interconnected porosity obtained by the Archimedes method.

Related to the samples dispersed with SA, the sedimentation behaviour is lessened. The porosity values reported are around 90 vol.% across the sample height. Therefore, these suspensions seem more stable despite not generating a complete steric effect as compared with KD4. The nanometric particles of NiO used in this research show a higher degree of dispersion in a non-polar media (camphene) when a short carbon-chain additive (SA) is employed. Stearic acid seems to reduce the effective size and weight of the particles, which allows an increase in $V_{cr}$ that enables particle pushing and redistribution during the camphene solidification. Furthermore, SA reduced the tendency to form agglomerates or molecule entanglements with PS or the solvent.

In samples with SA, the interconnected porosity obtained by Archimedes' method (Figure 4) was lower than the total porosity, meaning that close porosity was well-distributed across the sample height and properly detected by the measurement technique. In contrast, in the sample with KD4, the interconnected porosity was preferentially located from the middle to the top part of the sample, thereby showing the tendency of particle sedimentation effect in the samples with KD4.

According to Figure 5, a significant pore morphology modification across the sample height is observed when the cooling temperature and dispersant agent varies. The KD4 dispersant generates a greater pore size and morphology variation along with sample height than SA. Notably, the pore structure created during the directional solidification retains its many features after the thermal cycle of reduce-sintering. The variation in porosity along with height is promoted by the application of the directional solidification during the sample fabrication. The creation of a gradient pore-size structure is now visible, where fewer and smaller pores can be observed in the lower part of the samples, and pore size is then increased as the cooling rate diminishes. The gradient porosity is enhanced in samples with KD4 due to the particle sedimentation, especially at the sample 30KD4 (Figure 5). The number of small pores in the lower part of the sample is moderated as the cooling temperature increases.

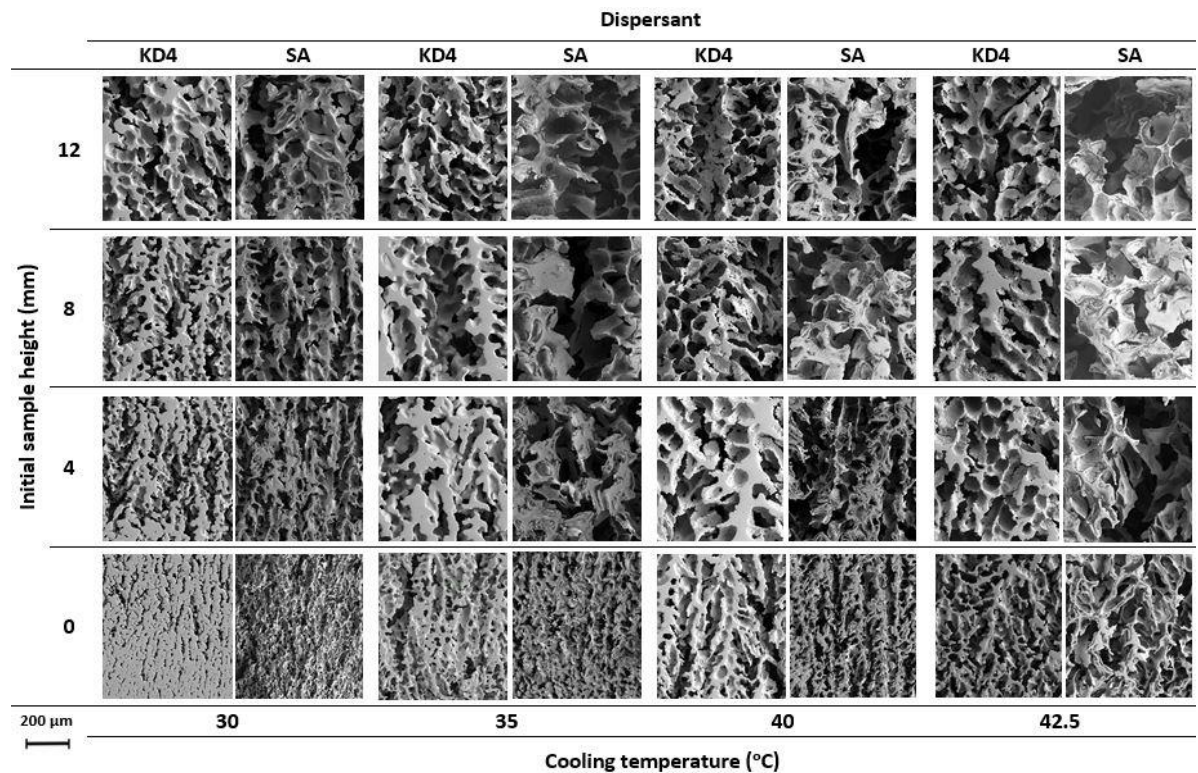

Figure 5. Effect of cooling temperatures (30, 35, 40, and 42.5 °C) and the dispersant agent (KD4 or SA) on the pore morphology of Ni wicks across the initial sample height. FESEM micrographs were taken using the same magnification (200x). Scale bar placed at the bottom left of the figure represents 200 μm.

The use of SA leads to larger pore sizes with less variation (gradient) according to the sample height. NiO powder dispersed with SA in camphene-based suspensions shows smaller effective particle size, as is demonstrated by the increase of viscosity with the milling time (Figure 3), and the particle concentration remains constant during solidification. While $V^h$ increases and $G^h$ decreases, the smaller NiO particles are pushed much further since the camphene crystal interface experiences a transition from cellular to equiaxed [48]. The dendrite contour is also more recognisable since the particle stacking and formation of the secondary arms have been promoted. Ultimately, larger pores have fewer probabilities to disappear during the reduction-sintering heat treatment. It should be emphasised that, after the sintering process under reducing conditions, the pore structure created during directional solidification retains its many features.

Cooling conditions show a clear influence not only on the pore size, but also on the pore morphology. Once the solidification starts, dendrite growth is driven by the high $G^h$ and low $V^h$ leading to an elongated crystal. During the course of solidification, $G^h$ is reduced and $V^h$ increased, and camphene dendrite turns gradually in to an equiaxed crystal since secondary and tertiary arms can grow more easily due to mitigation of the directional cooling and the lower possibility of particle entrapment [30]. This change in pore morphology is visible in Figure 5 and is correlated as a function of $G^h$ and $V^h$ in Figure 6. Figure 6 also shows the increment in the final pore size of the fabricated Ni wicks as a function of the cooling temperature (30, 35, 40, and 42.5 ºC) and the initial sample height ($h$ = 4, 8, and 12 mm), whereby both dispersants reveal a strong dependency of $G^h$ and $V^h$ on pore size and morphology obtained during the FC. Nevertheless, under similar cooling conditions, SA produced wider, more equiaxed pores.

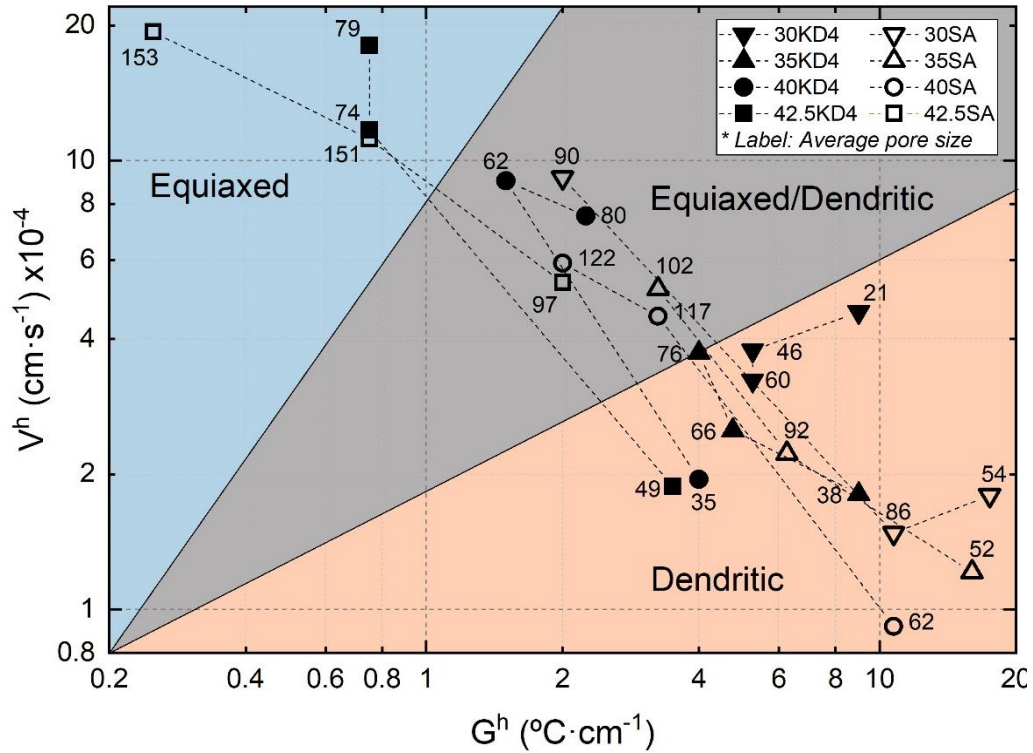

Figure 6. Average pore size and typical morphology features of the Ni wicks as a function of the average thermal gradient ($Gh$) and solidification front velocity ($Vh$) measured at different heights($h$ = 4, 8, and 12 mm) during sample solidification. Point labels indicate the average pore size in microns obtained at each cooling temperature (30, 35, 40, and 42.5 ºC) and dispersant employed (KD4 or SA). Equiaxed and dendritic refers to its characteristic microstructure.

Figure 7 shows the 3D reconstruction of the Ni wicks 30KD4 and 42.5KD4 in their lowest and topmost parts, respectively. In the lower part of sample 30KD4 (Figure 7a-b) with the highest $G^h$ and the lowest $V^h$ recorded, this cooling condition produced a highly interconnected structure of narrow porous channels aligned with the direction of the solidification, as can be seen in the pore skeleton render. On the other hand, in the upper part of the sample 42.5KD4 (Figure 7c-d), which showed the lowest $G^h$ and the highest $V^h$, a broadening of the pore size, and a transition from a dendritic and branched morphology to a more equiaxed morphology occurred due to the reduction of the number of pore connections. Indeed, as the solidification front moves through the sample, the directional driving force of the solidification diminishes due to the continuous reduction of $G^h$, as shown in Figure 6.

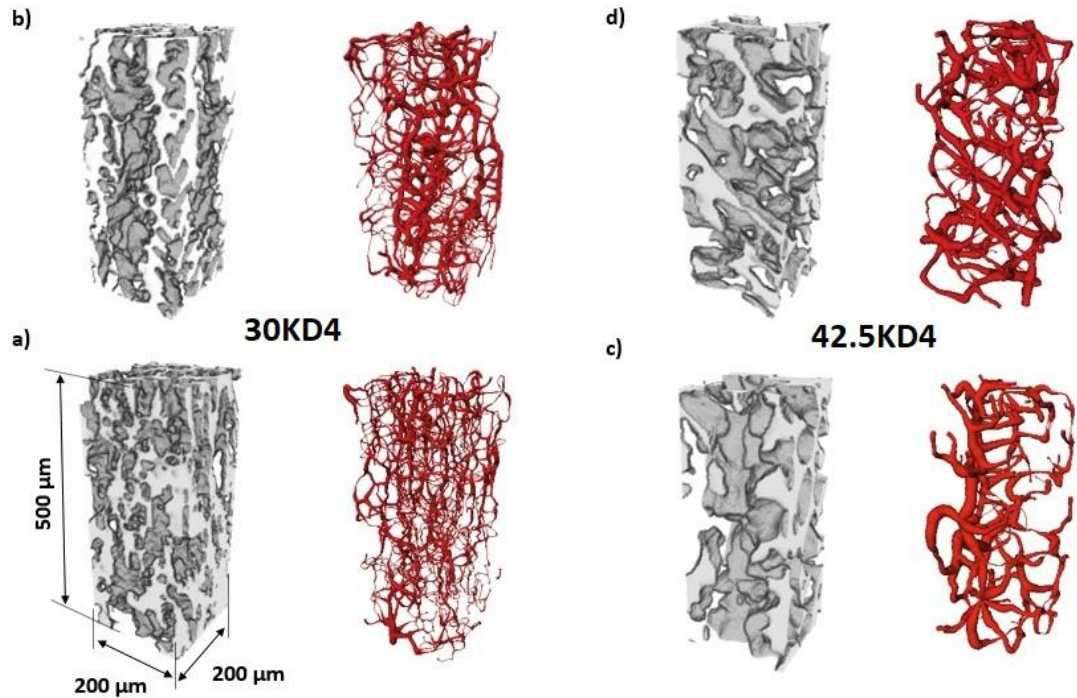

Figure 7. 3D reconstructions from X-ray computed tomography (X-CT) of Ni wicks fabricated by freeze casting using KD4 dispersant at two cooling conditions 30 ºC (a-b) and 42. 5 ºC (c-d). Samples (grey) and pore skeleton (red) were rendered at different initial sample heights: 0 (a), 4 mm (b) for sample 30KD4, 8 (c), and 12 (d) mm for sample 42.5KD4. All reconstructed parallelepipeds are approximately 200 x 200 x 500 μm in size

3.3 Capillary performance of fabricated wicks

Figure 8 shows the fraction of water mass absorbed due to the interconnected porosity versus time as a function of the dispersant and cooling temperature. Ni wicks manufactured with KD4 have shown a higher mass absorption than those with SA. They have presented a large-pore alignment (i.e., dendritic pore morphology is widely extended) with the direction of solidification parallel to the advancing liquid front. Sample 30KD4 (Figure 8) presents the lowest adsorption rate and 1failed to 2did not completely fill despite having a wide dendritic pore morphology and the smallest pore size. This behaviour could be attributed to the lack of interconnected porosity (Figure 4) produced by the particle sedimentation at the bottom of the wick (Figure 5). Samples 42.5, 40, and 35KD4 present an increased absorption rate, respectively, as they show a lower extent of the equiaxed pore morphology in their top sections (Figure 6). Wicks made from suspensions dispersed with SA (30-42.5SA) have similar absorption rates (Figure 8), and reach a maximum fraction of absorbed water mass of 0.35, due to their similar pore sizes and widely extended equiaxed pore morphology through the sample (Figure 6). This behaviour is enhanced on the 42.5SA sample, which, despite possessing the largest pore size, shows the lowest absorption rate due to its completely equiaxed pore structure (Figure 6). Despite the limitation of Lucas-Washburn's (LW) law [52], it has been successfully applied to model and characterise the capillary pumping performance [37–39] of wick samples. The behaviour predicted for porous materials fits to those with pores are oriented perpendicular to the fluid surface and that resemble straight cylinders, although several modifications, such as gravity, fluid evaporation, tortuosity, and pore shape, have been proposed for it to adequately fit a broad range of situations. In the case of Ni wicks fabricated by the FC technique, the adsorption rates fall under the predicted values, most likely due to the influence of the pore morphology. A fractal dimension for the tortuosity factor ($D_T$) has recently been proposed [53] as $\sim t^{1/2D_T}$ in a derivation of the LW law (e.g., LW law applied with $D_T = 1$), in order to consider the variety of pore shapes and interconnectivity levels. This factor has already been employed to correct the influence of the pore morphology in samples fabricated by FC [54], and validates the influence of dendritic pore morphology in the improvement of the ability of liquid to flow through a porous structure.

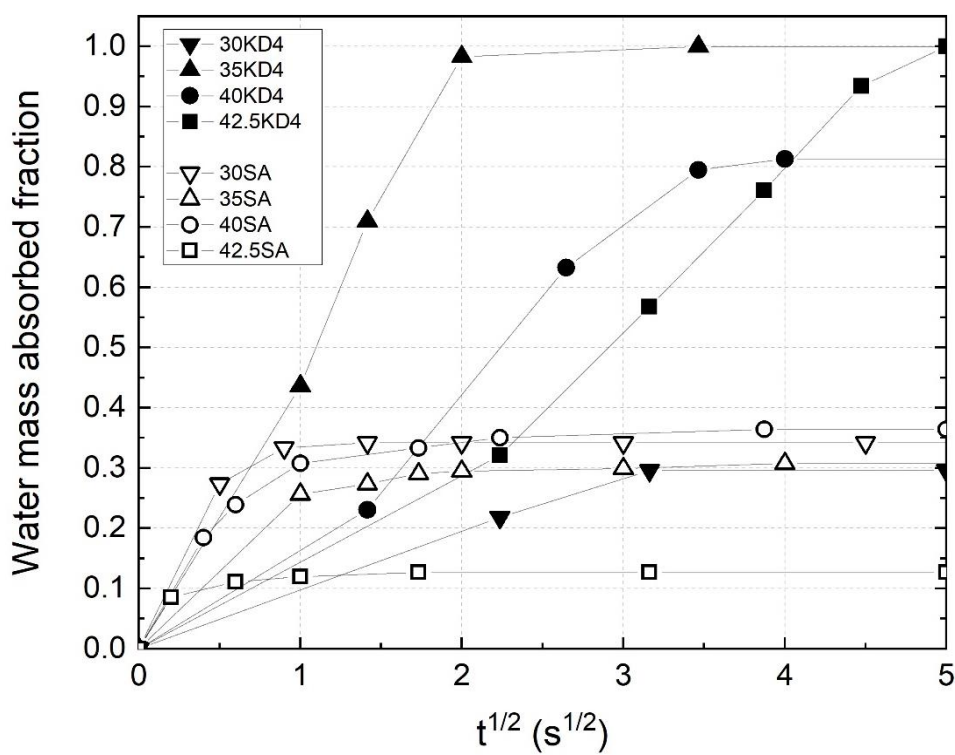

Figure 8. Water mass absorbed fraction over time of the Ni wicks as a function of the dispersant agent (KD4 or SA) and the cooling temperature (30, 35, 40, and 42.5 °C).

## 4. Conclusions

Ni wicks with gradient porosity were successfully fabricated by the FC technique. The processing modification implemented confirmed the effect of the dispersant agent on the final microstructure and on the wick performance. Stearic acid has been demonstrated to improve microstructural characteristics, such as dendrite contour definition (i.e., high tortuosity), and to increase the pore size. The use of dispersant with a short carbon-chain length creates a suspension of greater stability, reduces particle sedimentation thereby preserving the pouring ability. However, when SA is employed, the applied thermal gradient must be incremented to ensure a more directional dendritic pore structure, which constitutes the key factor for improved wick performance. In this regard, although KD4 shows higher particle sedimentation, it was able to promote (within the cooling rate employed) a more dendritic pore structure across the wick height.

The wick performance is linked to the dendritic pore size and the extension of the primary arm. The use of SA diminishes the particle sedimentation and enhances pore interconnectivity, although higher thermal gradients must be applied in order to increment the number of elongated camphene dendrites during solidification for the capillarity performance to be improved.

**Author Contributions:** Conceptualization, R.S.; methodology, R.S. and A.P.; formal analysis, R.S., A.P. P.LL.; investigation, R.S., P.LL.; data curation, R.S. and E.C.; writing—original draft preparation, P.LL.; writing—review and editing, R.S., E.C.; funding acquisition, R.S.. All authors have read and agreed to the published version of the manuscript.

**Funding:** Financial support for this work has been provided by the Spanish Ministerio de Economía, Industria y Competitividad (MINECO), through the project MAT2016-76713-P. Lloreda-Jurado P.J. thanks to the Universidad de Sevilla for the financial support (grant PIF II.2A, through VI Plan Propio de Investigación).

**Conflicts of Interest:** Declare conflicts of interest or state.